\documentclass{jfm}

\usepackage{graphicx, graphics}
\usepackage{natbib}

\ifCUPmtlplainloaded \else
 \checkfont{eurm10}
 \iffontfound
   \IfFileExists{upmath.sty}
     {\typeout{^^JFound AMS Euler Roman fonts on the system,
                  using the 'upmath' package.^^J}%
      \usepackage{upmath}}
     {\typeout{^^JFound AMS Euler Roman fonts on the system, but you
                  dont seem to have the}%
      \typeout{'upmath' package installed. JFM.cls can take advantage
                of these fonts,^^Jif you use 'upmath' package.^^J}%
     }
 \else
 \fi
\fi

\ifCUPmtlplainloaded \else
 \checkfont{msam10}
 \iffontfound
   \IfFileExists{amssymb.sty}
     {\typeout{^^JFound AMS Symbol fonts on the system, using the
               'amssymb' package.^^J}%
      \usepackage{amssymb}%
        \let\leq=\leqslant
        
     }{}
 \fi
\fi

\ifCUPmtlplainloaded \else
 \IfFileExists{amsbsy.sty}
   {\typeout{^^JFound the 'amsbsy' package on the system, using it.^^J}%
    \usepackage{amsbsy}}
   {\providecommand\boldsymbol[1]{\mbox{\boldmath $##1$}}}
\fi

\newsavebox{\astrutbox}
\sbox{\astrutbox}{\rule[-5pt]{0pt}{20pt}}

\newcommand\Xsin{\mbox{sin}}
\newcommand\Xcos{\mbox{cos}}
\newcommand\Xsec{\mbox{sec}}
\newcommand\Xlog{\mbox{ln}}

\title[Effective slip in  flow past super-hydrophobic stripes]{Effective slip in pressure-driven flow past super-hydrophobic stripes}

\author[A. V. Belyaev and O. I. Vinogradova ]%
{A.\ls V.\ls B\ls E\ls L\ls Y\ls A\ls E\ls V$^{1, 2}$
\ls \and
O.\ls I.\ls V\ls I\ls N\ls O\ls G\ls R\ls A\ls D\ls O\ls V\ls A$^{1, 2, 3}$
}

\affiliation{$^{1}$ Faculties of Physics and Chemistry, M. V. Lomonosov Moscow State University, 119991 Moscow, Russia \\[\affilskip]
$^2$A. N. Frumkin Institute of Physical Chemistry and Electrochemistry, Russian Academy of Sciences,
31 Leninsky Prospect, 119991 Moscow, Russia\\
[\affilskip]
 $^{3}$ ITMC and DWI,  RWTH Aachen, Pauwelsstr. 8, 52056 Aachen, Germany}
 \date{\today}
\label{firstpage}

\begin{document}

\maketitle

\begin{abstract}

Super-hydrophobic array of grooves containing trapped gas (stripes), have the potential to greatly reduce drag and enhance mixing phenomena in microfluidic devices. Recent work has focused on idealized cases of stick-perfect slip stripes. Here, we analyze the experimentally more relevant situation of a pressure-driven flow past striped slip-stick surfaces with arbitrary local slip at the gas sectors. We derive approximate formulas for maximal (longitudinal) and minimal (transverse) directional effective slip lengths, that are in a good agreement with the exact numerical solution for any surface slip fraction. By representing eigenvalues of the  slip length-tensor, they allow us to obtain the effective slip for any orientation of stripes with respect to the mean flow. Our results imply that flow past stripes is controlled by the ratio of the local slip length to texture size. In case of a large (compared to the texture period) slip at the gas areas, surface anisotropy leads to a tensorial effective slip, by attaining the values predicted earlier for a perfect local slip. Both effective slip lengths and anisotropy of the flow decrease when local slip becomes of the order of texture period. In the case of small slip, we predict simple surface-averaged, isotropic flows (independent of orientation).

\end{abstract}

\section{Introduction}

The development of microfluidics has motivated interest in
manipulating flows in very small channels~\citep{stone2004,squires2005}. Most of microfluidic devices operates with a pressure flow, which is faced with two main difficulties at this scale and under typical operating conditions. First, it is difficult to drive such a flow due to huge hydrodynamic resistance. Second, it is very difficult to mix, which normally requires a generation of a tranverse flow.

An efficient strategy for moving efficiently fluid in a tiny channel is to exploit hydrodynamic slip, which can be generated at hydrophobic surfaces and is quantified by the slip
length $b$ (the distance within the solid at which the flow profile extrapolates to zero)~\citep{vinogradova1999,lauga2005,bocquet2007}. Since for hydrophobic smooth and homogeneous surfaces $b$ can be of the order of tens of nanometers ~\citep{vinogradova2003,charlaix.e:2005,joly.l:2006,vinogradova.oi:2009}, but not much more, it is impossible to benefit of such a nanometric slip for pressure-driven microfluidic applications.
However, super-hydrophobic (SH) textures can significantly amplify hydrodynamic slip due to gas entrapment~\citep{vinogradova.oi:1995b,cottin_bizonne.c:2003.a}  leading to the huge slip length at the gas area. The composite nature of the texture, however, requires
regions of lower slip (or no slip) in direct contact with the liquid,
so the effective slip length of the surface $b_{\rm eff}$
is reduced. Indeed, experimental studies of flow past SH surfaces suggest
that effective slip is of the order of several microns ~\citep{ou2005,joseph.p:2006,choi.ch:2006}.

\begin{figure}
\begin{minipage}[b]{0.5\linewidth}
\centering
\includegraphics[height=3.7 cm, trim=0 -10 0 0]{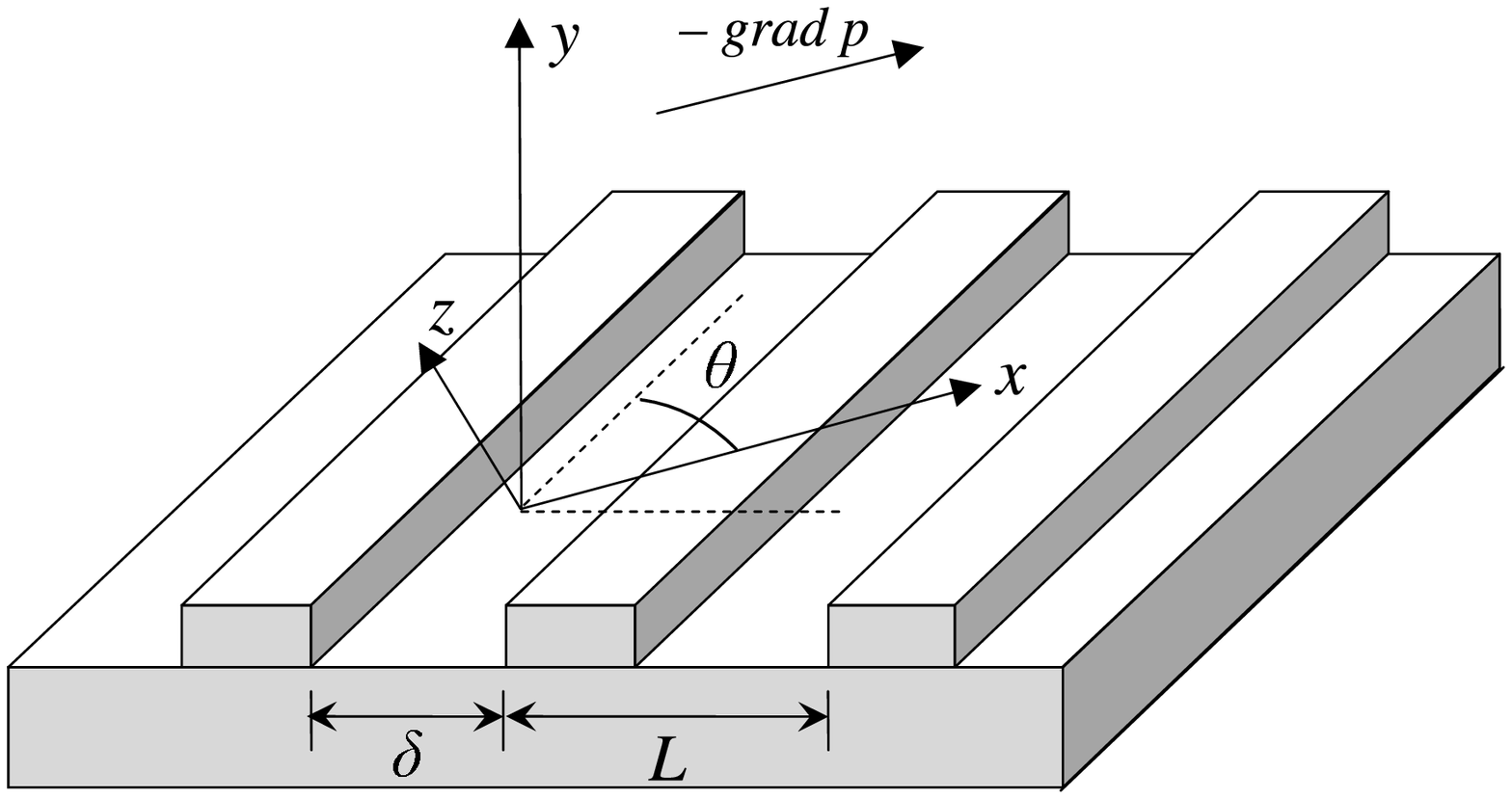}
\end{minipage}
\hspace{0.5cm}
\begin{minipage}[b]{0.5\linewidth}
\centering
\includegraphics[height=3.7 cm, trim=0 0 0 -10]{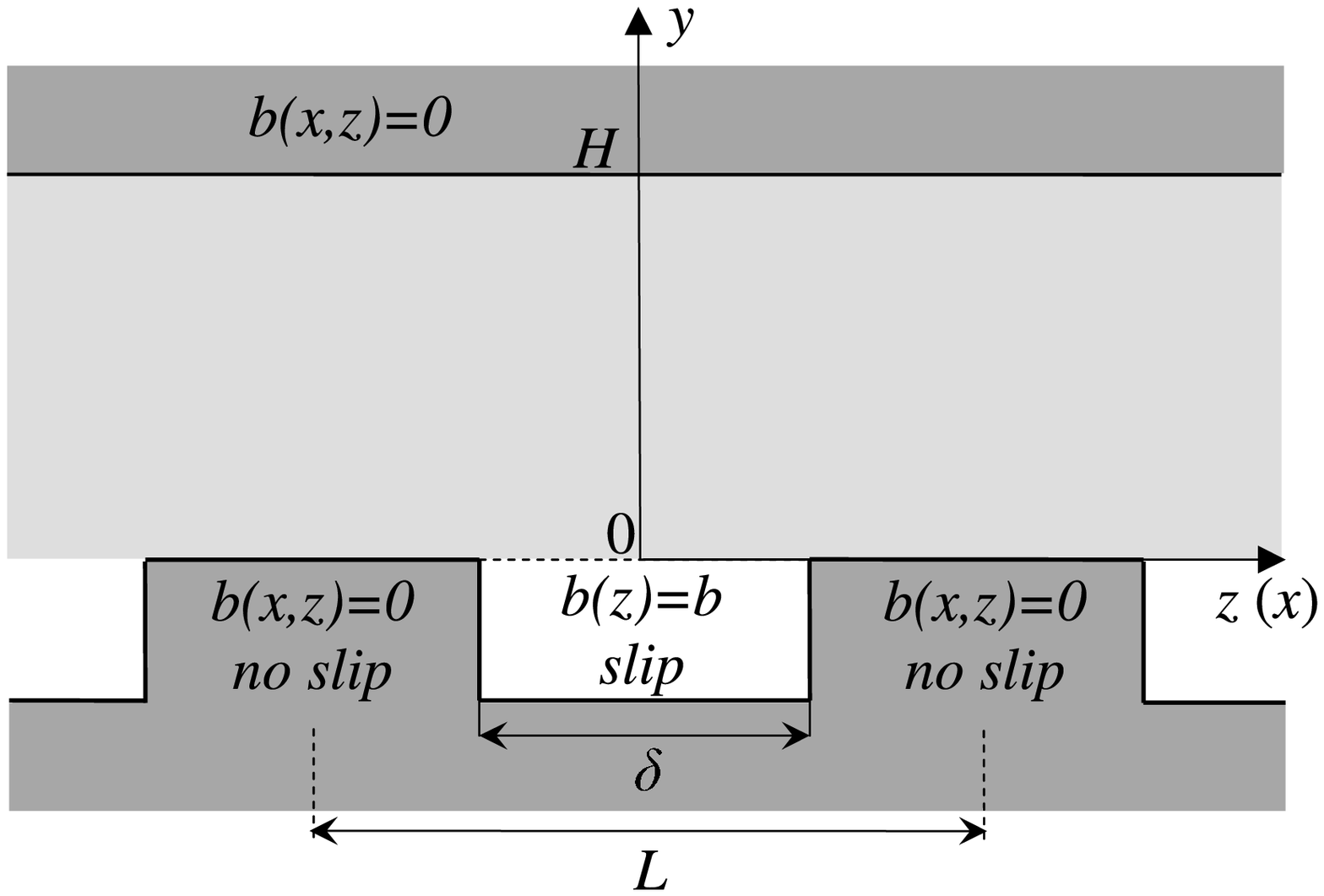}
\end{minipage}
\caption{(Left) Sketch of SH stripes: $\theta=\pi/2$ corresponds to transverse, whereas $\theta=0$ to longitudinal stripes; (right) situation in (left) is approximated by a periodic cell of size $L$, with equivalent flow boundary conditions on gas-liquid and solid-liquid interface. }\label{fig:geometry}
\end{figure}

SH surfaces consisting of periodic array of grooves containing trapped gas (Cassie's state)  are especially interesting since they allow to highlight effects of anisotropy. For anisotropic textures $b_{\rm eff}$ varies with the orientation of the wall texture relative to flow and is generally a tensor~\citep{Bazant08}. Such surfaces have been already used for reduction in pressure-driven flows~\citep{ou2005} and enhancement of mixing~\citep{ou.j:2007}. The problem of flow past stripes has been examined theoretically mostly with a typical geometry sketched in  Fig.~\ref{fig:geometry} corresponding to  of a roughly flat (no meniscus curvature) liquid interface, so that the modeled SH surface appeared as a perfectly smooth with a pattern of
boundary conditions. In the case of thin channels ($H \ll L$, where $H$ is the channel thickness, and $L$ is the period of the texture) the problem was solved for any two-component (e.g. low-slip and high-slip) texture, and striped surfaces were shown to provide rigorous upper and lower bounds on the effective slip over all possible two-phase patterns~\citep{feuillebois.f:2009}. The quantitative understanding of liquid slippage past such a surface in the thick channel ($H \gg L$) is however still challenging.  Pressure-driven flow has been analyzed for an idealized case of a perfect slip at the gas area \citep{philip.jr:1972,Lauga03,cottin.c:2004,sbragaglia.m:2007} and led to
\begin{equation}\label{effectiveslip}
b_{\rm eff}^\perp = \frac{L}{2 \pi }\Xlog\Big[\Xsec\left(\frac{\pi \phi_2}{2 } \right)\Big]
\ \ \ \mbox{ and } \ \ \ b_{\rm eff}^\| = 2 b_{\rm eff}^\perp,
\end{equation}
where $\phi_2=\delta/L$ denotes the fraction of the liquid-gas interface (correspondingly, $\phi_1=1-\phi_2$ is the fraction of solid-gas area), with the typical length scale of the slipping area $\delta$, and $b_{\rm eff}^\perp$ and $b_{\rm eff}^\|$ denote effective transverse and longitudinal slip lengths. Following ~\cite{Bazant08}, these are the eigenvalues of the second-rank effective slip-length tensor ${\bf b_{\rm eff}}$ represented by symmetric, positive definite $2\times 2$ matrix diagonalized by a rotation:
\begin{equation}  \label{effectiveBtensor}
{\bf b_{\rm eff}}= {\bf S}_{\theta} \left(
\begin{array}{cc}
 b^{\parallel}_{\rm eff} & 0\\
 0 & b^{\perp}_{\rm eff}
\end{array}
\right) {\bf S}_{-\theta} ,
\qquad
{\bf S}_{\theta}=
\left(
\begin{array}{cc}
 \Xcos~\theta &  \Xsin~\theta \\
 -\Xsin~\theta & \Xcos~\theta
\end{array}
\right). \label{Tensor_Rotation}
\end{equation}
Therefore, Eqs.~(\ref{effectiveslip}) allow to calculate $b_{\rm eff}$ in any direction given by an angle $\theta$ (Fig.~\ref{fig:geometry}).

Eqs.(\ref{effectiveslip}) provide an upper
limit for the effective slip lengths and in many situations would be expected to overestimate them. One reason is the possible meniscus curvature, which has been clarified in recent work~\citep{sbragaglia.m:2007, harting.j:2008, lauga2009}. Another is viscous dissipation  taking place in the underlying gas phase. Indeed, the more realistic ``gas cushion model''\citep{vinogradova.oi:1995a} predicts the finite slip length at the slipping area
\begin{equation}
b=e\left(\frac{\eta}{\eta_{\rm g}}- 1\right)\approx e \frac{\eta}{\eta_{\rm g}},
\end{equation}
where $e$ is the thickness of the gas layer, $\eta$ is the viscosity of liquid, and $\eta_{\rm g}$ is the viscosity of gas. Taking into account that under typical conditions $\eta/\eta_{\rm g}\approx 50,$ the variation of the SH texture height, $e$, in the typical interval $0.1-10$ $\mu$m \citep{quere.d:2005} gives $b=5-500$ $\mu$m, i.e. $b$ might be as small as typical $L$ or even less. For this reason, it is attractive to consider this experimentally relevant situation. However, despite its fundamental and practical significance, pressure-driven flow over partial slip stripes has received little attention. This has been studied numerically\citep{cottin.c:2004,priezjev.nv:2005,ybert.c:2007}. Nevertheless, no analytical resolution of the Stokes equation with this set of boundary conditions has been performed up to now.

In this paper, we provide analytical solutions to pressure-driven flows over SH stripes. In $\S2$ we formulate the problem and derive expressions for the effective slip for longitudinal and transverse stripes, which allows us to obtain a solution for any orientation of stripes with respect to a gradient of pressure. In $\S3$ we compare our results with numerical calculations performed by C. Cottin-Bizonne and C. Barentin using the method developed in \cite{cottin.c:2004} and discuss implications for the use of SH stripes to control hydrodynamic flows. We conclude in $\S4$.

\section{Model and Analysis}

We consider a pressure-driven flow past an idealized, flat, periodic, striped SH surface in the Cassie state (sketched in Fig.\ref{fig:geometry}), where the liquid-solid interface has no slip ($b_1=0$) and the liquid-gas interface has partial slip ($b_2=b$). Our results apply to a single surface in a thick channel ($H \gg \max\{L,b\}$), but not to thin channels ($H \ll \min\{ L, b \}$) where the effective slip scales with the channel width~\citep{feuillebois.f:2009}.
 The origin of coordinates is placed it the plane of liquid-gas interface above the middle of the slot. The $x$-axis is defined along the pressure gradient, while the $y$-axis is aligned across the channel. According to ~\cite{Bazant08} the general problem reduces to computing the two eigenvalues, $b_{\rm eff}^{\parallel}$ and $b_{\rm eff}^\perp$, which attain the maximal and minimal directional slip lengths, respectively.

The fluid flow satisfies Stokes' equations
\begin{equation}\label{NS}
   \eta\nabla^2\textbf{u}=\nabla p,\quad
  \nabla\cdot\textbf{u}=0,
\end{equation}
where $\textbf{u}$ is the velocity vector, and the applied pressure gradient is parallel to the $x$ axis direction:
\begin{equation}\label{PressGrad}
   \nabla p_0 = (-\sigma, 0, 0)
\end{equation}
The slip boundary conditions at the channel walls are defined in the usual way:
\begin{equation}\label{BC1}
   {\bf u}(x,0,z) = b(x,z)\cdot\frac{\partial {\bf u}}{\partial y}(x,0,z), \quad\hat{{\bf y}}\cdot {\bf u}(x,0,z) = 0.
\end{equation}
\begin{equation}\label{BC2}
   \textbf{u}(x,H,z)=-b_H\cdot\frac{\partial\textbf{u}}{\partial y}(x,H,z),\quad\hat{{\bf y}}\cdot {\bf u}(x,H,z) = 0.
\end{equation}
Here the local slip length $b(x,z)$ is generally the function of both $x$ and $z$ coordinates.
For simplicity, we now consider here the case $b_H=0$. As the problem is linear in \textbf{u}, we seek the solution in the form:
\begin{equation}\label{6}
   \textbf{u}=\textbf{u}_0+\textbf{u}_1,
\end{equation}
where $\textbf{u}_0$ is the velocity of the flow over the homogeneous plane with the no-slip condition:
\begin{equation}\label{7}
   \textbf{u}_0=(u_0, 0, 0),\qquad u_0=-\frac{\sigma}{2\eta} y^2 + C^*_0 y
\end{equation}
\begin{equation}
   C^*_0\equiv\frac{\partial u_0}{\partial y}(y=0)=\frac{\sigma H}{2\eta},
\end{equation}
and $\textbf{u}_1$ is the perturbation of the flow, which is caused by the presence of the texture and decays far from the bottom of the channel.

We are interested in the effective slip length $b_{\rm eff}$ of the lower surface defined as
\begin{equation}
   b_{\rm eff}=\frac{\langle u_s\rangle}{\langle \left(\frac{\partial u}{\partial y}\right)_s \rangle},
\end{equation}
where $\langle \ldots\rangle$ means the average value in plane $xOz$.

\subsection{Longitudinal stripes}

In this case the problem is homogeneous in $x$-direction ($\frac{\partial}{\partial x}=0$). The slip length $b(x,z)=b(z)$ is periodic in $z$ with period $L$. The elementary cell is determined as $b(z)=b$ at $|z|\leq\delta/2$, and $b(z)=0$ at $\delta/2<|z|\leq L$.
In this case velocity $\textbf{u}_1=(u_1, 0, 0)$ has only one nonzero component, which can be determined by solving the Laplace equation with the boundary conditions discussed above. By choosing $L/(2\pi)$ as the length scale and $\sigma L^2/(4\pi^2\eta)$ as the velocity scale we obtain in the dimensionless form
\begin{equation}\label{8}
   u_1(y,z)=\frac{a_0}{2}+\sum^\infty_{n=1} a_n \cos(n z) e^{-ny}.
\end{equation}
(The sine terms vanish due to symmetry.)
Condition (\ref{BC1}) leads to the dual trigonometric series
\begin{equation}\label{DoubleSeries_a}
   \frac{a_0}{2}+\sum^\infty_{n=1} a_n \left(1+\frac{2\pi b}{L} n \right) \cos(n z)=\frac{2\pi b}{L} C_0,\quad 0<z\leq c,
\end{equation}
\begin{equation}\label{DoubleSeries_b}
   \frac{a_0}{2}+\sum^\infty_{n=1} a_n \cos(n z)=0,\quad c<z\leq \pi,
\end{equation}
where
$c =\pi \phi_2$ and $C_0=C^*_0\cdot2\pi\eta/(\sigma L)=\pi H/L$.
To solve these series we assume that
\begin{equation}\label{10}
   \frac{a_0}{2}+\sum^\infty_{n=1} a_n \cos(n z)=\cos(z/2)\int\limits_z^c{\frac{h(t)dt}{\sqrt{\cos z -\cos t}}},\quad 0<z\leq c.
\end{equation}
According ~\citep{Sneddon} we then get
\begin{equation}\label{11}
   a_0=\frac{2}{\pi}\left[\frac{\pi}{\sqrt{2}}\int\limits_0^c{h(t)dt}\right],
\end{equation}
\begin{equation}\label{12}
   a_n=\frac{2}{\pi}\left[\frac{\pi}{\sqrt{2}}\int\limits_0^c{h(t)\left(P_n(\cos t)+P_{n-1}(\cos t)\right)dt}\right], \quad n=1,2,3,\ldots,
\end{equation}
where $P_n$ is Legendre polynomial, and one can then show that the effective slip length is given by
\begin{equation}\label{13}
   b_{\rm eff}^{\parallel}=\frac{L}{2 \pi}\frac{a_0}{2 C_0}.
\end{equation}
By integrating (\ref{DoubleSeries_a}) in the interval $[0,z]$, and substituting (\ref{10}) and (\ref{12}) we obtain ($0<z\leq c$)
\begin{equation}\label{14}
   \frac{2\pi b}{L} \int\limits_0^z{\frac{h(t)dt}{\sqrt {\cos t -\cos z}}}=\sec{\frac{z}{2}}\left[\frac{2\pi b}{L} C_0 z - \int\limits_0^z{\cos\left(\frac{\xi}{2}\right)\int\limits_\xi^c{\frac{h(t)dt}{\sqrt{\cos \xi -\cos t}}}d\xi}\right].
\end{equation}
We further change the order of integration in parentheses to get
\begin{equation}\label{change order}
   \int\limits_0^z{\cos\left(\frac{\xi}{2}\right)\int\limits_\xi^c{\frac{h(t)dt}{\sqrt{\cos \xi -\cos t}}}d\xi}=\int\limits_0^z{h(t)\int\limits_0^t{\frac{\cos\left(\frac{\xi}{2}\right)d\xi}{\sqrt{\cos \xi -\cos t}}}dt}+\int\limits_z^c{h(t)\int\limits_0^z{\frac{\cos\left(\frac{\xi}{2}\right)d\xi}{\sqrt{\cos \xi -\cos t}}}dt}
\end{equation}
The evaluation of (\ref{14}) gives
\begin{equation}\label{integral1}
   \int\limits_0^z{\frac{\cos\left(\frac{\xi}{2}\right)d\xi}{\sqrt{\cos \xi -\cos t}}}=\sqrt2\cdot\arcsin\left(\frac{\sin\frac{z}{2}}{\sin\frac{t}{2}}\right),
\end{equation}
so that we get
\begin{equation}\label{integral2}
   \frac{2\pi b}{L} \int\limits_0^z{\frac{h(t)dt}{\sqrt {\cos t -\cos z}}}=\sec\frac{z}{2}\left[\frac{2\pi b}{L} C_0 z-\frac{\pi a_0}{2}+\sqrt2\int\limits_z^c{h(t)\arccos\left(\frac{\sin\frac{z}{2}}{\sin\frac{t}{2}}\right)dt}\right]
\end{equation}
This can be simplified by neglecting the last term in parentheses, which is small as compared to the main term $\displaystyle \frac{\pi a_0}{2}$ (due to properties of $\arccos\left(\frac{\sin\frac{z}{2}}{\sin\frac{t}{2}}\right)$) and, thus,
\begin{equation}\label{h}
   h(t)=\frac{2}{\pi}\frac{d}{dt}\int\limits_0^t{\frac{\sin\frac{\xi}{2}}{\sqrt{\cos \xi -\cos t}}\left(C_0\xi - \frac{a_0 \pi}{2 \cdot \frac{2\pi b}{L}}\right)d\xi},
\end{equation}
whence
\begin{equation}\label{eqfora0}
   a_0=\frac{2\sqrt{2}}{\pi}\left[C_0\cdot \pi\sqrt2\ln\left(\sec\frac{c}{2}\right) - \frac{a_0 \pi}{2 \cdot \frac{2\pi b}{L}}\cdot\sqrt2\ln\left(\sec\frac{c}{2}+\tan\frac{c}{2}\right)\right].
\end{equation}
In what follows
\begin{equation}\label{beffpar}
   b_{\rm eff}^{\parallel}=\frac{L}{\pi} \frac{\ln\left[\sec\left(\displaystyle\frac{\pi \phi_2}{2 }\right)\right]}{1+\displaystyle\frac{L}{\pi b}\ln\left[\sec\displaystyle\left(\frac{\pi \phi_2}{2 }\right)+\tan\displaystyle\left(\frac{\pi \phi_2}{2}\right)\right]}.
\end{equation}

\begin{figure}
  \includegraphics[ width=6.5 cm, clip]{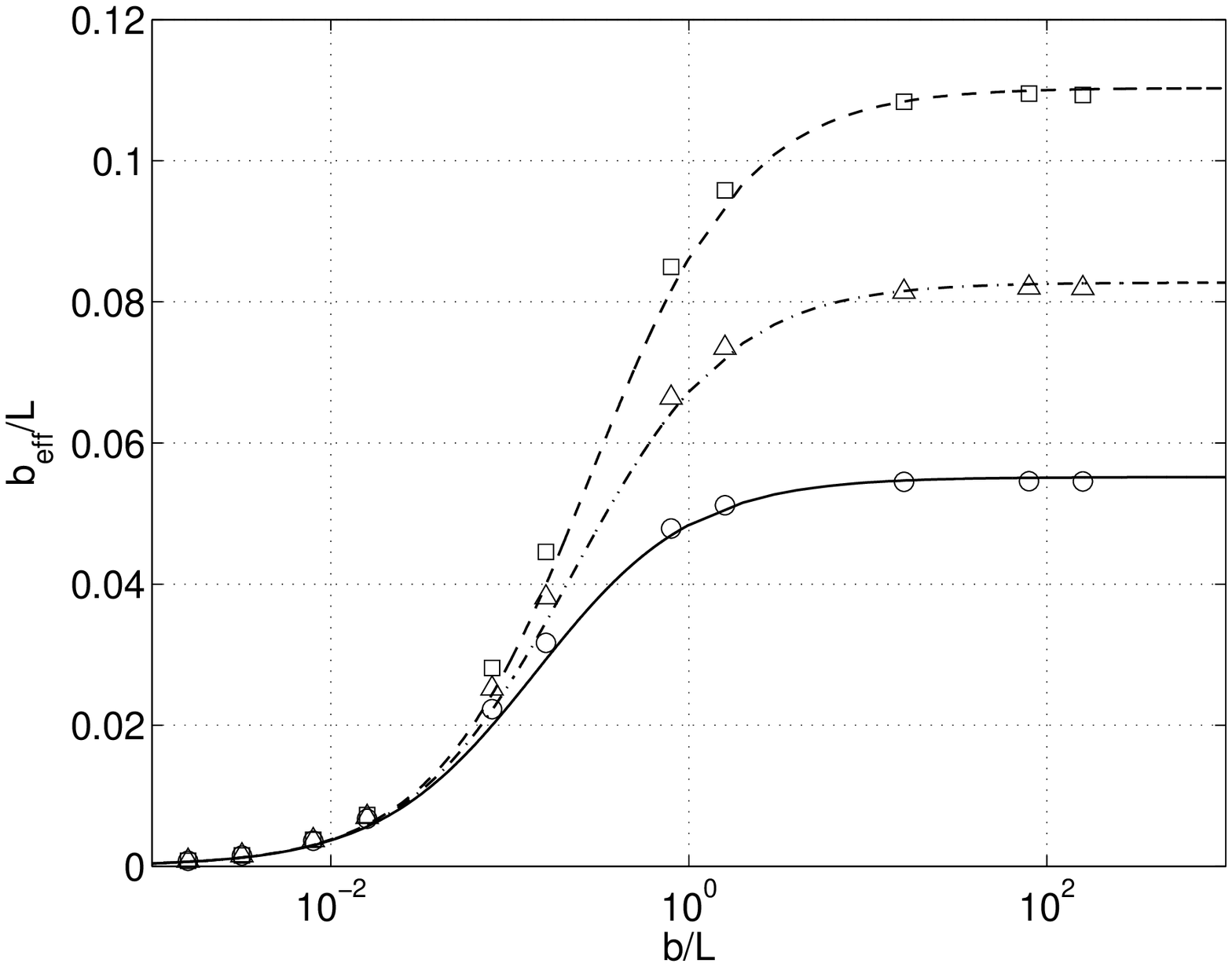}
  \includegraphics[width=6.5 cm, clip]{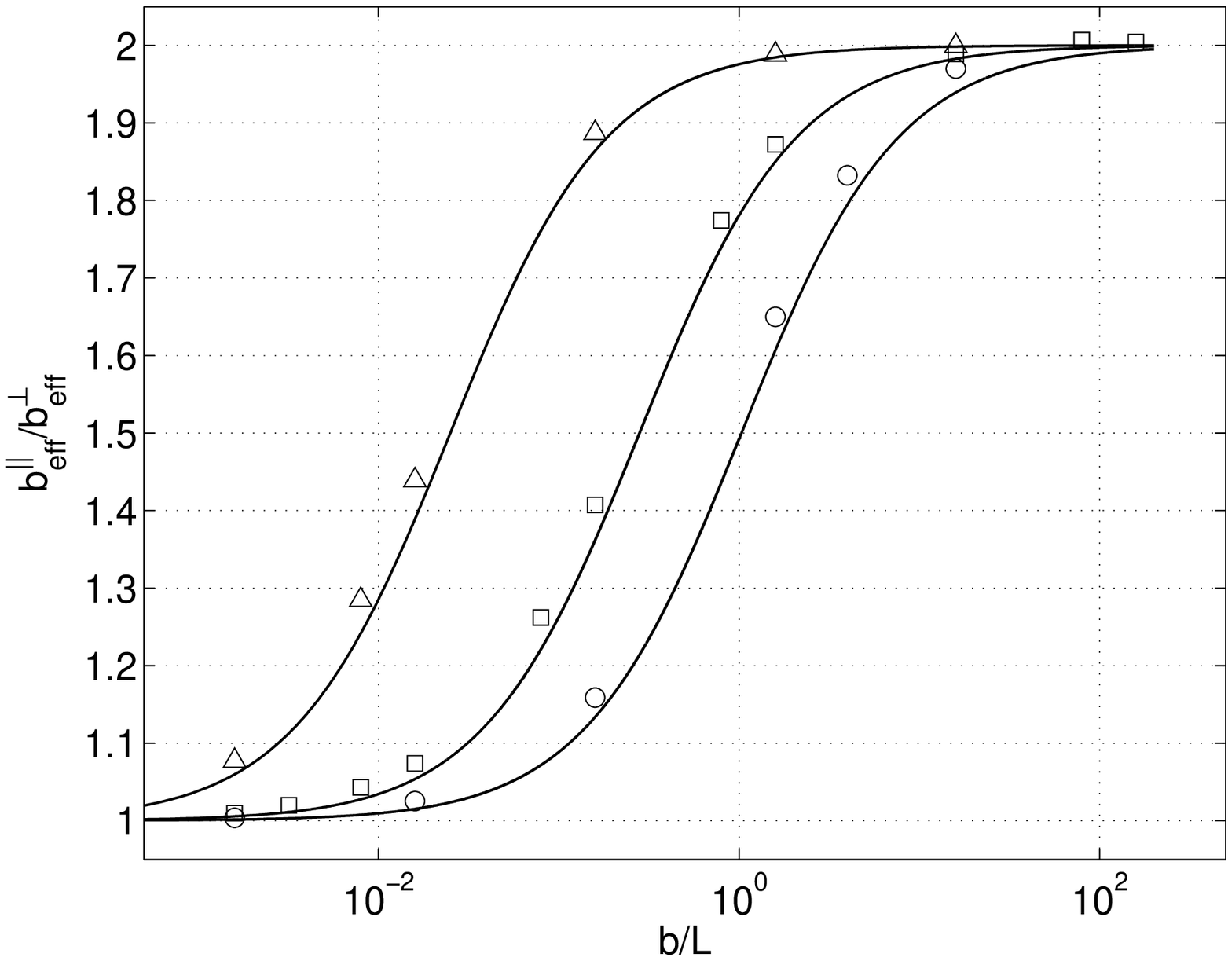}
  \caption{(Left) Eigenvalues $b_{\rm eff}^{\parallel}$ (dashed curve) and $b_{\rm eff}^{\perp}$ (solid curve) of the slip-length tensor ${\bf b_{\rm eff}}$ for stick-slip stripes of period $L$ and slipping area fraction $\phi_2=0.5$ as a function of the local slip length $b$ of this area. Dash-dotted curve represents the effective slip in the direction of driving force for tilted ($\theta=\pi/4$) stripes. Symbols show numerical results. (Right) The ratio of theoretically predicted eigenvalues of the slip-length tensor ${\bf b_{\rm eff}}$ (solid curves) and corresponding results of numerical modeling (symbols). From left to right, $\phi_2=0.05, 0.5, {\rm and} \, 0.95$.}
  \label{fig:b_sim}
\end{figure}

\subsection{Transverse stripes}

In this case the pressure gradient depends on $x$, so that it is convenient to introduce a stream function $\psi(x,y)$ and the vorticity vector $ \boldsymbol{\omega} (x,y) $. The two-dimensional velocity field corresponding to the transverse configuration is represented by $\textbf{u}(x,y)=\left(\frac{\partial \psi}{\partial y}, -\frac{\partial \psi}{\partial x} , 0 \right)$, and the vorticity vector $ \boldsymbol {\omega} (x,y)=\nabla\times\textbf{u}=(0,0,\omega)$ has only one nonzero component equal to
\begin{equation}\label{vorticity}
   \omega=-\nabla^2\psi.
\end{equation}
The solution can then be presented as the sum of the base flow with homogeneous no-slip condition and its perturbation due to the presence of stripes
\begin{equation}\label{SF01}
   \psi=\Psi_0+\psi_1,\quad \omega=\Omega_0+\omega_1,
\end{equation}
where $\Psi_0$ and $\Omega_0$ correspond to a typical Poiseuille flow
\begin{equation}\label{Psi0}
   \Psi_0=-\frac{\sigma}{\eta}\frac{y^3}{6}+C^*_0\frac{y^2}{2},\quad \Omega_0=\frac{\sigma}{\eta}y-C^*_0.
\end{equation}

The problem for perturbations $\psi_1$ and $\omega_1$ of the stream function and $z$-component of the vorticity vector reads
\begin{equation}\label{eq_psi1}
   \nabla^2\psi_1=-\omega_1,\quad \nabla^2\omega_1=0,
\end{equation}
which can be solved by applying boundary conditions (\ref{BC1}) and (\ref{BC2}), that take form \citep{priezjev.nv:2005}:
\begin{equation}\label{BC1_new}
   \frac{\partial \psi_1}{\partial y}(x, y=0)= b(x)\cdot\left[C^*_0-\omega_1(x, y=0)\right],
\end{equation}
\begin{equation}\label{BC2_new}
   \frac{\partial \psi_1}{\partial y}(x, y=H)= 0,
\end{equation}
and an extra condition that reflects our definition of the stream function
\begin{equation}\label{BC_psi}
   \psi_1(x,y=0) = 0.
\end{equation}

This can be solved exactly to get
\begin{equation}\label{GenSolOmega1}
   \omega_1(x,y) = \frac{\alpha_0}{2}+\sum^\infty_{n=1} \alpha_n \cos(\lambda_n x) e^{-\lambda_n y},
\end{equation}
\begin{equation}\label{GenSolPsi1}
   \psi_1(x,y) = -\frac{\alpha_0}{4}y^2+\beta_0 y + \sum^\infty_{n=1} \left(\beta_n + \frac{\alpha_n}{2} \frac{y}{\lambda_n} \right) \cos(\lambda_n x) e^{-\lambda_n y},
\end{equation}
where $\lambda_n=(2\pi n)/L$ is the wave-number. Condition (\ref{BC2_new}) leads to $\beta_0=\alpha_0 H/2$, and (\ref{BC_psi}) gives $\beta_n=0$.

 Applying boundary conditions, we obtain another dual series, similar to (\ref{DoubleSeries_a}) and (\ref{DoubleSeries_b})
\begin{equation}\label{DoubleSeries_c}
   a_0 \left(1+\frac b H \right)+\sum^\infty_{n=1} a_n \left(1+2 \cdot \frac{2\pi b}{L} n \right) \cos(n x)=\frac{2\pi b}{L} C_0 ,\quad 0<x\leq c,
\end{equation}
\begin{equation}\label{DoubleSeries_d}
   a_0 +\sum^\infty_{n=1} a_n \cos(n x)=0,\quad c<x\leq \pi.
\end{equation}
Here
\begin{equation}
   a_0=\frac{4 \pi^2 \eta }{\sigma L^2}\beta_0, \quad a_n=\frac{\alpha_n}{2 n}\frac{2\pi\eta}{\sigma L},    
\end{equation}
and $b_{\rm eff}^{\perp}=(L/2\pi)(a_0/C_0)$. Since $ b/H $ is negligibly small, the dual series can be simplified to obtain
\begin{equation}\label{beffperp}
b_{\rm eff}^{\perp}=\frac{L}{2 \pi} \frac{\ln\left[\sec\left(\displaystyle\frac{\pi \phi_2}{2 }\right)\right]}{1+\displaystyle\frac{L}{2 \pi b}\ln\left[\sec\displaystyle\left(\frac{\pi \phi_2}{2 }\right)+\tan\displaystyle\left(\frac{\pi \phi_2}{2}\right)\right]}
\end{equation}

\subsection{Arbitrary direction}

Here we consider the situation when the pressure gradient is aligned at the angle $\theta$ to the stripes. The surface velocity $\textbf{u}_s=(u_s, 0, w_s)$ has only two nonzero components. We establish the coordinate system so that $-\nabla p_0$ is parallel to the $x$-axis. According to \cite{Bazant08}
\begin{equation}\label{tensorial_BC}
   \langle\textbf{u}_s\rangle=\bf{b_{\rm eff}}\cdot\left\langle\left(\frac{\partial \textbf{u}}{\partial y}\right)_s\right\rangle,
\end{equation}
where $\bf{b_{\rm eff}}$ is given by Eq.(\ref{effectiveBtensor}).
%
Average components of surface velocity then read
\begin{equation}\label{aver_us}
  \langle u_s\rangle=(b_{\rm eff}^{\parallel}\cos^2\theta+b_{\rm eff}^{\perp}\sin^2\theta)\cdot C_0^*,
\end{equation}
\begin{equation}\label{aver_ws}
  \langle w_s\rangle=(b_{\rm eff}^{\parallel}-b_{\rm eff}^{\perp})\sin\theta\cos\theta\cdot C_0^*.
\end{equation}

The absolute value of the slip velocity on the striped SH-surface $|\textbf{U}_s|$ and the angle $\varphi$ between the driving force $(-\nabla p_0)$ and $\textbf{U}_s$ are then given by
\begin{equation}\label{slip_velocity}
  |\textbf{U}_s|=\frac{\sigma H}{2 \eta} \sqrt{ (b_{\rm eff}^{\parallel}\cos\theta)^2+(b_{\rm eff}^{\perp}\sin\theta)^2},\: \tan\varphi=\frac{(b_{\rm eff}^{\parallel}-b_{\rm eff}^{\perp})\sin\theta\cos\theta}{(b_{\rm eff}^{\parallel}\cos^2\theta+b_{\rm eff}^{\perp}\sin^2\theta)}.
\end{equation}

\section{Discussion}

Figure~\ref{fig:b_sim}~(left) shows the  theoretical eigenvalues of the slip-length tensor ${\bf b_{\rm eff}}$ for a slipping area fraction $\phi_2=0.5$ as a function of the slip length $b$ calculated with Eqs.~(\ref{beffpar}) and (\ref{beffperp}). In addition, we plot the data for tilted stripes ($\theta=\pi/4$). Also included in Figure~\ref{fig:b_sim}~(left) are results of a numerical solution of the Stokes' equations performed by C. Cottin-Bizonne and C. Barentin using the method developed in \cite{cottin.c:2004}. The agreement between a theory and simulation data is very good
for all $\phi_2$ and $b/L$, but at $b/L=O(1)$ there is
some small discrepancy, suggesting that our formulas slightly underestimate the effective slip, which is likely due to a simplification of Eq.~(\ref{integral2}). The same trends were observed for other values of $\phi_2$. Still, our analytical expressions for the effective slip, Eqs.~(\ref{beffpar}) and (\ref{beffperp}), appear to be surprisingly accurate, especially taking into account their simplicity. The same remark concerns the use of tensorial formula, Eq.(\ref{effectiveBtensor}).

Our results imply that flow past stripes is controlled by the ratio of the local slip length $b$ to texture period $L$. At $b/L \gg 1$ our expressions for $b_{\rm eff}$ turn to Eqs.~(\ref{effectiveslip}) suggested earlier for a perfect local slip. As expected, the effective slip decreases when $b/L=O(1)$ and smaller. Interestingly, this ratio also controls the anisotropy of the flow. Indeed,
combining Eqs.~(\ref{beffpar}) and (\ref{beffperp}) we get
\begin{equation}\label{ratio}
b_{\rm eff}^{\parallel}=b_{\rm eff}^{\perp} \left(1+ \frac{1}{1+\displaystyle\frac{L}{\pi b} \ln\left[\sec\displaystyle\left(\frac{\pi \phi_2}{2}\right)+\tan\displaystyle\left(\frac{\pi \phi_2}{2 }\right)\right]} \right)
\end{equation}
 If $b/L \gg 1$, the effective slip for parallel stripes, $b_{\rm eff}^{\parallel}$, is twice that of perpendicular stripes, $b_{\rm eff}^{\perp}$, as it was in case of a perfect slip ($b_2=\infty$) at the liqud-gas interface \citep{Lauga03,cottin.c:2004,sbragaglia.m:2007,bahga:2009}. In this case surface anisotropy leads to a truly tensorial effective slip. However, anisotropy of the flow decreases with a decrease in $b/L$, and at small $b/L$ we get $b_{\rm eff}^{\parallel, \perp}\sim b$. In other words, at small local slip we predict simple surface-averaged, isotropic flows (independent of orientation), which means despite the fact that the local slip varies in only one direction, the effective slip is scalar. These unexpected results are summarized in Fig. ~\ref{fig:b_sim} (right). This finding can be understood by using the following simple arguments. Following the advice of H.A.~Stone (private communication), let us consider the average fluid velocity $\langle u_s \rangle$ on  the SH surface. According to boundary condition (\ref{BC1})
\begin{equation}\label{average_u_s}
   \langle u_s \rangle = \frac{1}{L^2} \int\limits_0^L{\int\limits_0^L{u_s (x,z) dx}dz} = \frac{1}{L^2} \int\limits_0^L{\int\limits_0^L{b(x,z)\left(\frac{\partial u}{\partial y}\right)_s dx}dz}
\end{equation}
For transverse flow this expression takes the form
\begin{equation}\label{average_u_s_transv}
   \langle u_s \rangle = \frac{1}{L} \int\limits_0^\delta{b\left[C_0^*+ \left(\frac{\partial u_1}{\partial y}\right)_s\right] dx} = b C_0^* \phi_2 + \frac{b}{L} \int\limits_0^\delta{\left(\frac{\partial u_1}{\partial y}\right)_s dx},
\end{equation}
where $C_0^*=\left(\frac{\partial u_0}{\partial y}\right)_s= {\rm const}$ is, obviously, independent of the relative orientation of stripes in respect to a pressure gradient, since $u_0$ represents the solution of the problem for a smooth homogeneous surface. The same arguments apply in longitudinal case, where the only difference would be the integration over $z$ instead of $x$. Therefore, when $b$ is a small value $( b/L=O(\varepsilon) )$, the second term in (\ref{average_u_s_transv}) may be neglected as an infinitely small value of higher (second) order because $u_1\varpropto \varepsilon $,
and, thus,
\begin{equation}\label{beff_smallb}
   \left(b_{\rm eff}\right)_{b\rightarrow0} \approx b \phi_2 + O\left(\varepsilon^2\right)
\end{equation}
is independent of an external force direction. The anisotropy of the effective slip is determined by the second integral term in (\ref{average_u_s_transv}), which dominates when $b/L=O(1)$ and larger. These results suggest that both the value (upper limit) of the effective slip length and anisotropy of the flow are controlled by the smallest characteristic length of the problem (in our case, $b$ or $\delta$).

\begin{figure}
  \includegraphics[ width=6.5 cm, clip]{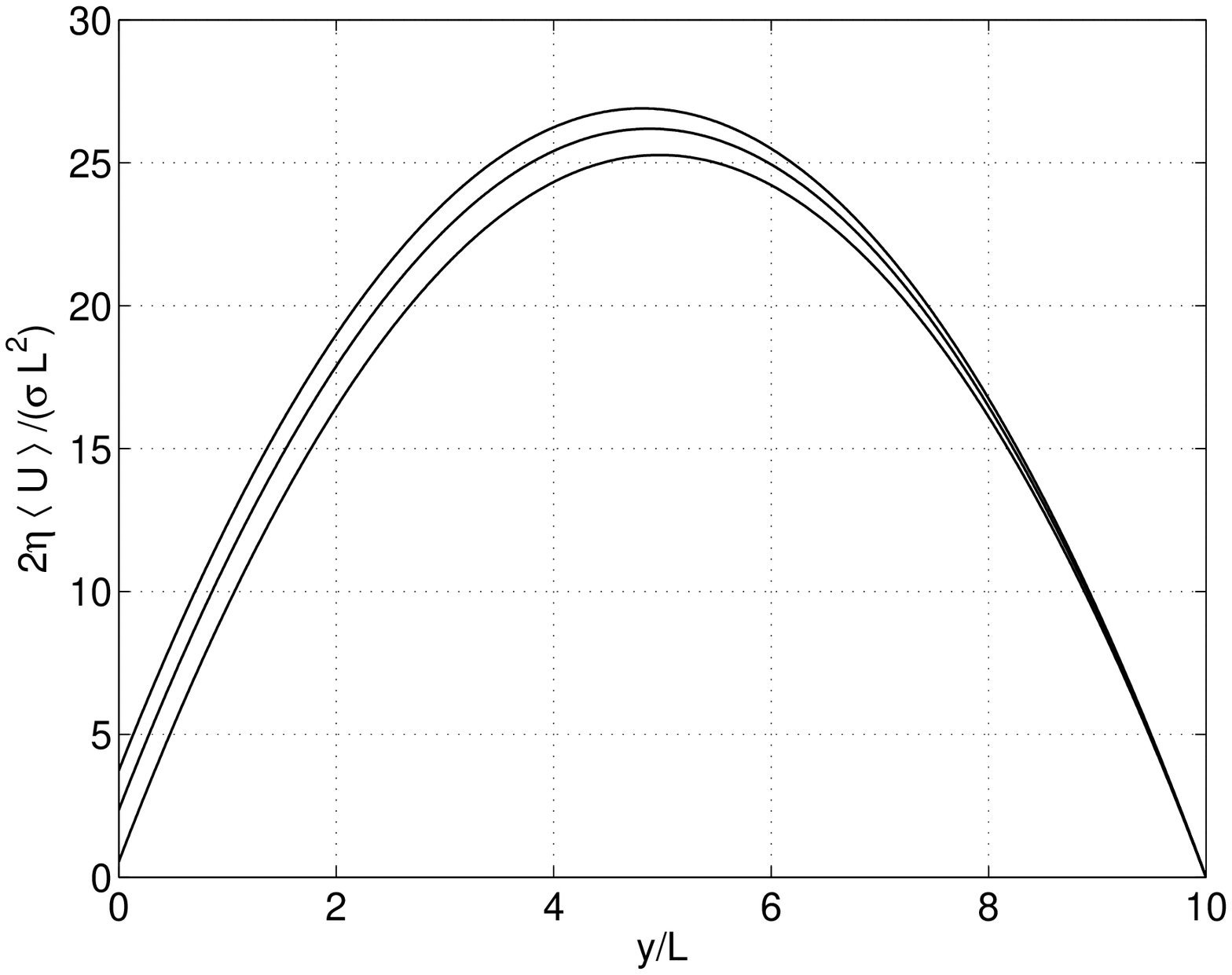}
  \includegraphics[width=6.5 cm, clip]{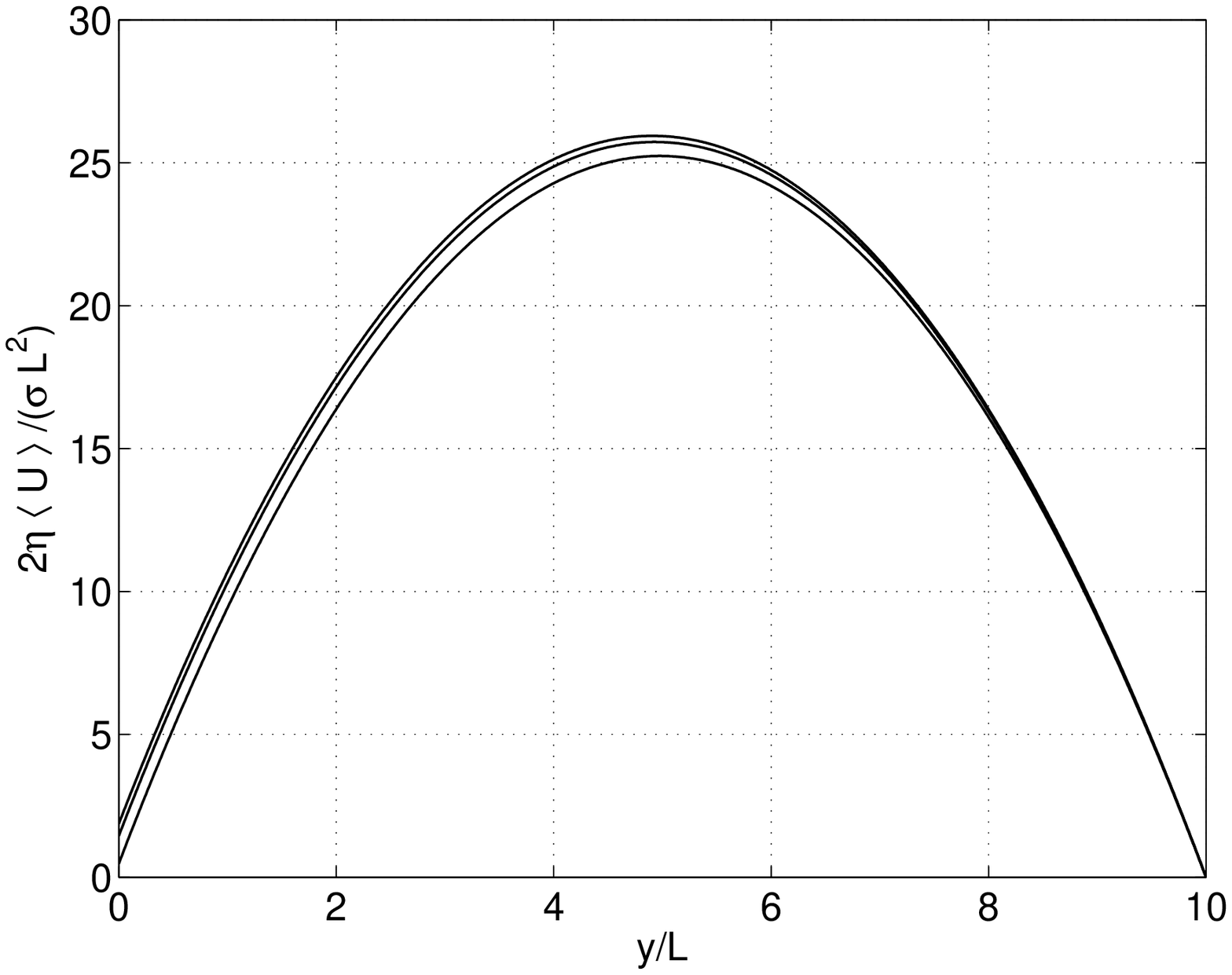}
  \caption{ Average velocity profiles ($\phi_2=0.8$) for a longitudinal (left) and transverse (right) flow. From top to bottom $b/L=1000$, $1$ and $0.1$}
  \label{fig:profile}
\end{figure}

Finally, we present average velocity profiles in longitudinal ($\theta=0$) and transverse ($\theta=\pi/2$) configurations for different values of local slip length $b$ (Fig.~\ref{fig:profile}). Mean flow remains two-dimensional and parabolic when the driving force is applied in main directions, yet both the average slip velocity at $y=0$ and the maximal velocity value at the middle of the channel depend on $b$. For arbitrary $\theta$ the flow is essentially three-dimensional as the orthogonal velocity component appears due to the tensorial effective boundary condition Eq.(\ref{tensorial_BC}).

\section{Conclusion}\label{sec:conclusion}

We have analyzed pressure-driven flow over striped SH surfaces. Unlike the previous approach, we have obtained general analytical solutions for any value of local partial slip. We have confirmed that the hydrodynamic response of a striped slipping surface is generally anisotropic. Our main conclusion is that both effective slip and flow anisotropy are controlled by the ratio of local slip at the gas area to texture size. When this ratio is large, our results are closely related to those of \cite{Lauga03,cottin.c:2004,sbragaglia.m:2007,bahga:2009}, and surface anisotropy leads to anisotropy of effective slip. For a small ratio we predict not only a decrease in the effective slip, but also a different, isotropic response of the striped SH surface.

\section*{Acknowledgement}
The advice of H.A.~Stone is gratefully acknowledged. We thank C.~Cottin-Bizonne and C.~Barentin for numerical verification of the theory. This research was supported by the DFG under the Priority programme ``Micro and nanofluidics'' (grant Vi 243/1-3) and by the RAS under the
Priority Program ``Assembly and Investigation of Macromolecular Structures of New Generations''.

\bibliographystyle{jfm}

\bibliography{jfm_Belyaev}

\end{document}